\documentclass{article}

\usepackage{arxiv}

\usepackage[utf8]{inputenc} 
\usepackage[T1]{fontenc}    
\usepackage{hyperref}       
\usepackage{url}            
\usepackage{booktabs}       
\usepackage{amsfonts}       
\usepackage{nicefrac}       
\usepackage{microtype}      
\usepackage{lipsum}
\usepackage{graphicx}
\graphicspath{ {./images/} }
\usepackage{amsmath}
\usepackage{algorithm}
\usepackage{algpseudocode}
\usepackage{xcolor} 

\title{Verified Tool Calls Improve LLM Agent Reliability Under
Non-Atomic Failures}

\author{
 Isham Kalappurackal Mansoor \\
  Department of Computer Science\\
  Old Dominion University\\
  Norfolk, VA 23529 \\
  \texttt{ikala001@odu.edu} \\
   \And
 Abhishek Phadke \\
  School of Engineering and Computing\\
  Christopher Newport University\\
  Newport News, VA 23606 \\
  \texttt{abhishek.phadke@cnu.edu} \\
  \And
 Pratip Rana \\
  Department of Computer Science\\
  Old Dominion University\\
  Norfolk, VA 23529 \\
  \texttt{prana@odu.edu} \\
}

\begin{document}
\maketitle
\begin{abstract}
Large Language Model (LLM) agents rely on external tools to perform multistage tasks. Existing agent frameworks typically assume that tool calls are atomic and return binary success or failure signals. However, real-world systems exhibit non-atomic behaviors such as timeouts after dispatch, delayed visibility, and partial state updates. These mismatches lead to reliability issues including duplicate actions, task success, and unnecessary tool executions. A lightweight, verification-aware tool wrapper is introduced that augments tool calls with postcondition verification, verify-before-retry logic, and idempotency keys. The approach is evaluated in a controlled simulated environment with injected non-atomic failures across multiple task templates. The results demonstrate that the proposed method significantly reduces duplicate actions, while maintaining comparable task success rates. Overall, the findings suggest that strengthening tool interaction semantics is a promising direction for improving LLM agent reliability without requiring modifications to the underlying language model.
\end{abstract}


\section{Introduction}

The deployment of Large Language Models (LLMs) as autonomous agents marks a transition from passive text generation to active environmental orchestration. These agents increasingly function as controllers over application programming interfaces (APIs), databases, and multi-step workflows, leveraging reasoning to bridge the gap between human intent and system execution \cite{vuddanti2026recoverabilitylawerrmeasure} \cite{11081716}. These agents typically rely on tool responses to guide their decision-making. However, most existing systems assume that tool calls are atomic: an operation either succeeds or fails \cite{yao2023reactsynergizingreasoningacting, anthropic2024mcp}. There exist many different paradigms like ReAct (Thought, Action, Observation) and protocols like MCP (Model Context Protocol) which focus on the preliminary mechanics of tool invocation, optimizing the processes by which AI models identify and assign parameters to available tools \cite{yao2023reactsynergizingreasoningacting, anthropic2024mcp}. However, these methods struggle with the execution realities of latency and out-of-sync states. These gaps highlight a fundamental limitation: while existing systems improve how agents decide to call tools, they largely overlook how to ensure those calls are reliably executed in flawed conditions. In real-world distributed systems, this assumption is frequently violated by behaviors that are not atomic, such as network timeouts occurring after a request has been dispatched, eventual consistency leading to delayed visibility of updates, and stale-version conflicts in multi-agent environments \cite{atomix2026, hellandpatidempotence, paul2025multiagent, Vogels2009EventuallyConsistent}. This mismatch leads to critical reliability issues, including duplicate actions, incorrect task completion, and cascading failures in long workflows. This information asymmetry causes the response channel (what the agent sees) and the effect channel (what actually happened on the server) to be separate \cite{yao2023reactsynergizingreasoningacting}. Therefore, a lightweight verification-aware wrapper is proposed that augments tool calls with postcondition checks and safe retry strategies.

\begin{figure}[h]
    \centering
    \includegraphics[width=0.8\linewidth]{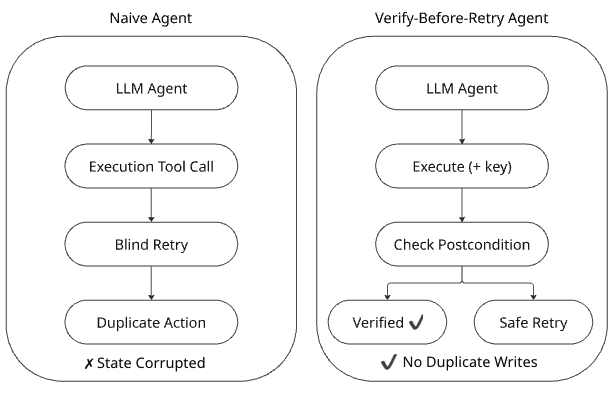}
    \caption{Diagram showing a verification-aware strategy compared to the naive approach.}
    \label{fig:placeholder}
\end{figure}

Naive retry strategies in LLM agents can lead to duplicate actions and corrupted state under failures such as timeouts. Our verification-aware wrapper addresses these issues by introducing postcondition validation and failure-aware retry strategies, enabling agents to reason not just about tool selection, but about the correctness of their effects in the external environment.

\textbf{Our Contributions:}
\begin{itemize}
\item Formalize a non-atomic tool failure as a distinct reliability problem in LLM agents, with a taxonomy of 4 failure modes.
\item Propose a verify-before-retry wrapper that queries postcondition state before retrying, preventing duplicate actions.
\item Show improved reliability without modifying the underlying LLM model, through this verify-before-retry wrapper framework.
\end{itemize}

\section{Background \& Problem Formulation}

\subsection{How LLM Agent Tool Calls are Modeled}

The modern LLM agent architecture traces back to the ReAct framework, which established that LLM agents should interleave reasoning traces and task-specific actions at each step, allowing the model to maintain, update, and adjust plans based on a selected state of the environmental feedback \cite{yao2023reactsynergizingreasoningacting} \cite{qu2025comprehensivereviewaiagents}.


A tool call is formalized as a tuple $(a, r, \mathcal{S}, \mathcal{S}')$, where: 

\begin{itemize}
    \item $a \in \mathcal{A}$ is the action dispatched by the agent
    \item $r$ is the response returned to the agent by the tool
    \item $\mathcal{S}$ is the ground truth world state before the dispatch
    \item $\mathcal{S}'$ is the ground truth world state after the dispatch
\end{itemize}

Schick et al. (2024) showed that LLMs can be trained to decide which APIs to call, when to call them, what arguments to add, and how to incorporate results into future reasoning \cite{schick2023toolformerlanguagemodelsteach}. However, this work treated $r$ as a faithful oracle: it assumes that the response $r$ accurately reflects the resulting state $S'$. This assumption, however, does not hold in many real-world systems.

A tool is defined as atomic if and only if:

\[ r \equiv (S \rightarrow S') \]

This means that the response $r$ faithfully encodes the state transition from $\mathcal{S}$ to $\mathcal{S}'$: the action either visibly succeeds or visibly fails, with no ambiguity. This definition is consistent with the notion of atomicity in database systems, where an ACID transaction either commits completely or rolls back without exposing partial effects \cite{Ribeiro2015DataModeling}. By contrast, distributed systems often relax this guarantee to improve availability, making delayed visibility and other non-atomic behaviors common in practice \cite{Vogels2009EventuallyConsistent}.

\subsection{Why Real Tools Are Non-Atomic}

Non-atomicity is the emergent property of real production infrastructure. Three well-understood phenomena are causing it:

\begin{enumerate}
    \item \textbf{Network latency and API Timeouts}: When an agent dispatches an action $a$, the request travels over a network. If the timeout expires before a response arrives, the agent receives an error code, but the action may have already executed at the server \cite{huang2026agentsfailactdiagnostic}. It has been established that the main categories of error taxonomy for tool invocation failures cover parameter initialization, execution, and result interpretation, noting that systematic evaluation of tool use reliability in production settings remains underdeveloped \cite{huang2026agentsfailactdiagnostic}. This case is among the hardest to handle precisely because the agent has no reliable signals about whether execution occurred. 

    \item \textbf{Eventual Consistency}: Many production APIs, such as cloud databases, CRMs, and notification systems, do not provide strong read-after-write guarantees. Eventually consistent guarantees that if no new updates are made to an object, eventually all accesses will return the last updated value, but reads immediately after a write may return stale data \cite{Vogels2009EventuallyConsistent}. An agent that verifies success immediately after dispatch may query a replica that has not yet received the update, triggering a false negative verification and unnecessary retry. 

    \item \textbf{Partial Execution}: Compound actions may execute only partially due to internal service errors, partial batch failures, or transactional isolation mismatches. The returning status code may report success even though only a subset of intended effects were applied. 
\end{enumerate}

Failures in multi-agent systems are frequently complex, involving convoluted agent interaction and the compounding effects of individual model behaviors, and pinpointing the origin of failure requires understanding not just whether the system failed \cite{cemri2025multiagentllmsystemsfail}. Our work narrows this observation to a specific and underexplored failure mechanism: the agent's belief state diverges from ground truth because tool responses are not atomic, even when both the agent and the tool are individually correct.  

\subsection{Four Failure Modes of Non-Atomic Tools}

Most papers on LLM agent reliability focus on planning failures, hallucinations, or tool selection errors. Current approaches treat tool calling as a black box process where the LLM decides based on internal reasoning, without explicit mechanisms to verify whether preconditions for tool invocation are satisfied or whether the tool's output meets expected postconditions, this can lead to tools being called with incorrect parameters, invalid results being incorporated into reasoning, and the agent's representation of world state becoming inconsistent \cite{liu2026toolgatecontractgroundedverifiedtool}.

The wrapper framework is tested in 4 main failure modes that are common in most distributed systems:

\begin{enumerate}
    \item \textbf{Timeout-After-Dispatch:} A network timeout occurs after the server has begun processing (or completed) the action, but before the response reaches the agent.

    \item \textbf{Delayed Visibility:} The action successfully transitions the system from $\mathcal{S}$ to $\mathcal{S}'$, but a verification read observes stale state due to eventual consistency, returning $\mathcal{S}$ instead of $\mathcal{S}'$.

    \item \textbf{Partial Success:} The action is a compound operation. An internal service error causes only a subset of effects to be applied. The API may return HTTP 200 with partial data, or may surface no error at all. 


    \item \textbf{Stale Conflicts:} The action fails because another agent, thread, or process has modified the relevant state since the agent's last observation. The action was valid against the observed $\mathcal{S}$, but $\mathcal{S}$ has since become stale due to a concurrent write. 
\end{enumerate}

\section{Related Work}

\subsection{LLM Agent Reliability}

Recent work on LLM agents focuses on enabling models to interact with external tools through iterative reasoning and action loops. The ReAct framework introduced a paradigm where agents alternate between reasoning steps and tool calls, using tool outputs to guide subsequent decisions \cite{yao2023reactsynergizingreasoningacting}. Similarly, Toolformer trains models to decide when and how to invoke tools in a self-supervised manner \cite{schick2023toolformerlanguagemodelsteach}. 

A common assumption across these approaches is that tool calls behave reliably. When an agent executes an action, the returned response is treated as an accurate reflection of what happened in the environment. In other words, the response is assumed to represent the effect of the action faithfully. Benchmarks such as ToolBench\cite{xu2023toolmanipulationcapabilityopensource} and AgentBench\cite{ICLR2024_e9df36b2} evaluate how well agents complete tasks using tools, but they operate under the same assumption of clean, deterministic tool behavior. This introduces many limitations; for instance, these frameworks do not account for cases where the tool response is missing, delayed, or inconsistent with the system's actual state. As a result, they implicitly treat tool calls as atomic, which could break down in real-world deployments. 

\subsection{Retry and Error Handling in Agents}

Another line of work explores improving agent reliability through reflection and retry strategies. Methods such as Reflexion\cite{shinn2023reflexionlanguageagentsverbal} and Language Agent Tree Search\cite{zhou2024languageagenttreesearch} allow agents to revise their behavior after failures by incorporating feedback or exploring alternative action sequences. However, these approaches are only effective when failures are due to poor reasoning or incorrect planning. For example, if an agent selects the wrong tools or uses incorrect parameters, retrying with a revised plan can improve performance. Moreover, they do not address a different class of failures where the action itself may have already succeeded, but the agent does not know it. In such cases, retry is not a correction; it is a duplication. For instance, if a payment request succeeds but the response is lost, a retry will execute the same action again, leading to unintended side effects. 

\subsection{Distributed Systems Foundations}

There are well known problems in distributed systems that are similar to the issues faced by current agentic systems. In particular, unreliable communications between components can lead to situations where an operation is executed, but its acknowledgment is lost or delayed. This has led to the distinction between "at-least-once" and "exactly-once" execution semantics. In "at-least-once" systems, operations may be retried and therefore executed multiple times \cite{exactlyonce1997}. In contrast, "exactly-once" systems ensure that each operation is applied only once, even in the presence of failures\cite{exactlyonce1997}. A common practical solution is the use of idempotency keys, which allow systems to safely handle retries by ensuring that repeated requests with the same key do not produce duplicate effects.

Additionally, Eventually Consistent Systems highlight that in many real systems, updates are not immediately visible \cite{Vogels2009EventuallyConsistent}. This means that even if an action succeeds, reading the system state immediately afterward may not reflect the update, leading to incorrect conclusions about the failure. These phenomena directly correspond to the failure modes identified earlier, including lost responses, delayed visibility, and partial updates. While distributed systems provide well-established solutions to these problems, such as idempotency keys and consistency models, these ideas have not been systematically incorporated into LLM agent architectures. 



\subsection{Positioning of this Work}

This work addresses a gap at the intersection of LLM agents and distributed systems. Prior approaches focus either on improving the agent’s reasoning or on evaluating task performance under ideal conditions. In contrast, this framework focuses on the reliability of interactions between agents and tools. The main contribution is a verify-before-retry framework that explicitly checks whether an action has already succeeded before attempting it again. By combining postcondition verification with idempotent execution, this approach prevents duplicate actions and improves robustness under ambiguous tool responses. Importantly, this method operates at the system level and does not require changes to the underlying language model. It can therefore be integrated with existing agent frameworks while addressing a class of failures that prior work does not consider.

\section{Method: Verified Tool Wrapper }

\subsection{Design Principles}

The core failure modes identified in Section 2.3 share a common structural property: they arise from the divergence between the effect channel, the causal pathway through which a tool action modifies world state, and the response channel, the pathway through which the agent receives acknowledgment of that modification. In reliable distributed systems, these two channels are decoupled by design; an action may be dispatched, take effect, and yet produce no observable response, or produce a response that arrives out of order or is lost entirely \cite{Vogels2009EventuallyConsistent}. A naive agent that conflates "no response" with "no effect" will retry indiscriminately, risking duplicate writes, constraint violations, and inconsistent downstream state.

Our design is organized around three principles derived from this observation. First, effect response separation: the system must independently verify whether an action's postcondition holds in world state, without relying on the response channel. Second, verify-before-retry: the system must never issue a retry without first consulting a postcondition verifier, ensuring that retries are issued only when the intended state change is confirmed absent. Third, idempotent execution: when a retry is warranted, it must be issued with a key that enables server-side deduplication, so that a retry arriving after a delayed success does not produce a duplicate effect. Together, these principles constitute a strict discipline analogous to the at-least-once delivery with idempotent consumers pattern in distributed messaging, applied here to the LLM agent tool calling context \cite{hellandpatidempotence}.

\subsection{Task Suite}

To test this framework, two multistage workflow tasks are used. The goal is not to test whether the model can solve difficult reasoning problems, but whether it can safely complete ordinary tool workflows when tool feedback is unreliable.

\begin{itemize}
    \item \textbf{activate\_customer:} The agent must create a customer record, set the customer status to \texttt{ACTIVE}, and send exactly one welcome message. This task mainly tests duplicate side effects, since retrying a message send after a timeout can create more than one message.

    \item \textbf{record\_invoice:} The agent must update an invoice row, create a corresponding record, and set the record status to \texttt{RECORDED}. This task mainly tests partial writes, since a row or record may exist while still missing required fields.
\end{itemize}

These two tasks cover two important forms of non-atomicity. In the first task, the danger is doing the same external action too many times. In the second task, the danger is assuming an action succeeded just because some object exists. For this reason, our verifiers check the actual task postconditions rather than only checking object existence.

\subsection{Task Success Definition}

This section formalizes what "success" means when using this framework. Let $S'$ be the world state after action $a$ has been dispatched. The postcondition $C(S')$ is defined as a boolean predicate over $S'$:

\[C(S'): S \rightarrow \left\{0, 1\right\}\]

Where $C(S') = 1$ if and only if  all the expected effects of action $a$ are fully and correctly reflected in $S'$. The action $a$ is successful if and only if $C(S') = 1$. 
\\
\\

For the activate customer task, the postcondition would be:

\[
C(S') = \left[ \text{customer.status} = \text{"activate"} \right] 
\land 
\left[ \text{customer.activated\_at} \neq \text{null} \right]
\]

And for the invoice task, it would be:

\[
C(S') = \left[ \text{row.exists} \right] 
\land 
\left[ \text{row.invoice\_id = a.invoice\_id} \right]
\land
\left[
\text{row.amount = a.amount}
\right]
\]

Three main implications should be noted with these formalizations:

\begin{enumerate}
    \item \textbf{Verifier completeness is required, not existence checking}. A verifier that only checks $[row.exists]$ will pass on partial success (Failure Mode 3). So the system must follow all parts of the formalizations

    \item \textbf{Postconditions encode domain knowledge}. System cannot automatically infer $C(S')$ from an API call alone. This is a feature (the system is generalizable across domains) and a limitation (adoption requires engineering effort per task)

    \item \textbf{Postcondition accuracy affects the ablation}. If verifier produces false negatives on delayed visibility (Failure Mode 2), the verify-only variant will over suppress. 
\end{enumerate}

\subsection{Postcondition Verifier}

For each tool action $a$, a corresponding postcondition verifier $V_a$ is defined, which determines whether the intended effect of the action has been achieved in the current state. The verifier returns one of three outcomes:

\begin{itemize}
    \item \textbf{True}: the postcondition is satisfied.
    \item \textbf{False}: the postcondition is not satisfied.
    \item \textbf{Unknown}: the state is inconclusive.
\end{itemize}

These three valuded logic is necessary to handle eventual consistency, where state updates may not be immediately observable. Moreover, in practice, verifiers can be implemented in several ways:
\begin{itemize}
    \item \textbf{Direct State Reads:} Such as querying a database or API to check whether the desired state change has occurred.
    \item \textbf{API Polling:} Repeatedly querying until the system converges to a consistent state.
    \item \textbf{LLM-based interpretation:} When the state is unstructured, an LLM can be used to evaluate whether the postcondition holds.
\end{itemize}

A critical property is that verification is read only. Unlike tool actions, which modify state, verifiers only observe it. This ensures that repeated verification calls are safe and cannot introduce additional side effects. Moreover, the correctness of the verifier directly affects the system's behavior. For instance, incomplete verifiers may misclassify partial success as complete success. Moreover, overly strict verifiers may incorrectly trigger retries in eventually consistent systems.

\subsection{Idempotency Key Protocol}

To ensure safe retries, each tool invocation is associated with an idempotency key $k$, constructed as a deterministic function of the action:

\[ k = hash(\text{agent\_id, action\_type, payload, timestamp\_bucket}) \]

This key is passed with every execution attempt of the same logical operation. If the underlying system supports idempotency, repeated requests with the same key are guaranteed to produce at most one effect. This mechanism is widely used in distributed systems to approximate "exactly-once" execution in the presence of unreliable communication. 



When the underlying API does not support idempotency keys, the verifier still protects against duplicate effects. In this case, retries are only issued when the verifier confirms that the action has not yet succeeded. However, without idempotency, there remains a small window where concurrent retries or race conditions may still produce duplicates. Thus, idempotency keys provide stronger guarantees, while verification provides best effort protection.




\subsection{Model and Agent Implementation}
A single language model drove all agents, Google Gemini Flash-Lite, accessed through the langchain-google-genai interface. The same model was used for every method so that differences in the results reflect the tool interaction layer rather than differences in model capability. The decoding was configured with temperature $0.0$ and a maximum of $1024$ output tokens to keep the runs as deterministic as the provider allows. The agent loop itself is implemented as a ReAct style loop in LangGraph: the model alternates between an LLM node that selects tool calls and a tool node that executes them against the simulated environment.
 
Each method is defined entirely by its system prompt and the wrapper logic around the tool node; the underlying model, tools, and decoding parameters are held fixed. The baseline prompt instructs the agent to retry a failed tool call directly without checking external state and to omit idempotency keys. In all variants, the tool-call retry budget is capped at $N = 1$ retry per logical operation ($\texttt{MAX\_TOOL\_RETRIES\_PER\_CALL} = 1$); the verify-before-retry wrapper only issues that retry when the postcondition verifier reports that the action did not complete. Ambiguous intermediate states are resolved inside the wrapper and are not surfaced to the model.
 
For the main comparison, $25$ independent episodes were run per method in every configuration, giving $2\text{ tasks} \times 3\text{ fault levels} \times 2\text{ methods} \times 25 = 300$ agent runs in total. The ablation was run as a separate configuration, testing only the \textbf{activate\_customer} task on medium level. Due to a different run configuration, its absolute success and duplicate rates are incomparable to those in the Figures ~\ref{fig:main-comparison} and ~\ref{fig:table_main}. Fault injection is seeded deterministically (\texttt{base\_seed} $= 42$, incremented per episode), and the same seed sequence is reused across methods so that every method faces an identical stream of injected timeouts, delayed reads, partial writes, and conflicts. The LLM rate-limited responses are retried up to five times inside the client.

\subsection{Verify-Before-Retry Algorithm}






Overall, the algorithm ensures that retries occur only when necessary, and are safe when executed. 

\begin{algorithm}[h]
\caption{Verified Tool Call}
\begin{algorithmic}[h]

\Require action $a$, verifier $V$, idempotency key $k$, max retries $N$
\Ensure success or failure

\State execute($a$, key=$k$)

\For{$i = 1$ to $N$}

    \State response $\gets$ get\_response($a$)

    \If{response == SUCCESS}
        \State \Return success
    \EndIf

    \If{response == FAILURE (definitive)}
        \State \Return failure
    \EndIf

    \If{response == AMBIGUOUS}
    
        \State wait(backoff($i$))
        \State result $\gets V(\text{current\_state})$
        
        \If{result == TRUE}
            \State \Return success 
        \ElsIf{result == UNKNOWN}
            \State \textbf{continue} 
        \ElsIf{result == FALSE}
            \State execute($a$, key=$k$) 
        \EndIf
        
    \EndIf

\EndFor

\State \Return failure

\end{algorithmic}
\end{algorithm}

\section{Experimental Setup}

\subsection{Simulated Tool Environment}

The system is tested in a controlled simulated tool environment instead of using live external APIs. This gives full control over when and how failures occur, while still preserving the key problem. This setting follows the action observation loop used by tool using agents such as the ReAct framework \cite{yao2023reactsynergizingreasoningacting}, but adds an explicit external world state so that comparing what the agent believes happened with what actually happened is possible. The setup also allows us to inject faults using a fixed random seed, following the general idea of fault injection for dependability testing \cite{hsueh1997fault}. However, our focus is specifically on failures that matter for LLM agents using tools: actions that may complete, partially complete, or become temporarily invisible even though the agent receives an unreliable observation.

\subsection{Compared Methods}

For the ablation study, three variants were compared:

\begin{itemize}
    \item \textbf{Retry Only Baseline:} The agent retries directly after a timeout or tool error without checking the external state.

    \item \textbf{Verification Only:} The agent checks whether the intended postcondition already holds after an ambiguous failure. If the action is verified as complete, it does not retry. If the action is not verified, it reports failure without retrying.

    \item \textbf{Verify-before-retry:} The agent first verifies the postcondition after an ambiguous failure and only retries when the verifier indicates that the action did not complete.
\end{itemize}

This distinction is important because retrying can help when an action truly failed, but it can also create duplicate side effects when the action already succeeded.

\subsection{Failure Injection Protocol}

\begin{table}[h]
\centering
\begin{tabular}{lcccc}
\toprule
\textbf{Fault level} & \textbf{Timeout} & \textbf{Delayed visibility} & \textbf{Partial success} & \textbf{Conflict} \\
\midrule
Low    & 0.05 & 0.05 & 0.03 & 0.02 \\
Medium & 0.20 & 0.15 & 0.10 & 0.05 \\
High   & 0.35 & 0.25 & 0.15 & 0.10 \\
\bottomrule
\end{tabular}
\caption{Fault probabilities used in the simulated tool environment.}
\label{tab:fault-configs}
\end{table}

Each run used a fixed seed, 42, and the same seed is used across methods. As shown in Table ~\ref{tab:fault-configs}, there are 3 level, each level controls the probability of four fault types: timeout after dispatch, delayed visibility, partial success, and conflict. 

\subsection{Metrics}

Each method was evaluated using five metrics:

\begin{itemize}
    \item \textbf{Task Success Rate:} whether the final world state satisfies all required task postconditions.
    \item \textbf{Duplicate Action Rate:} whether the agent produced repeated side effects for an action that should happen once.
\end{itemize}

For success, duplicate action, and task success rates, proportions across repeated runs are reported. Where appropriate, Wilson score confidence intervals \cite{wilson1927probable} and standardized effect sizes using Cohen's $h$ \cite{cohen1988statistical} are used.

\section{Results}

\subsection{Reliability Gains Grow as Faults Worsen}

\begin{figure}[h]
\centering
\includegraphics[width=0.9\linewidth]{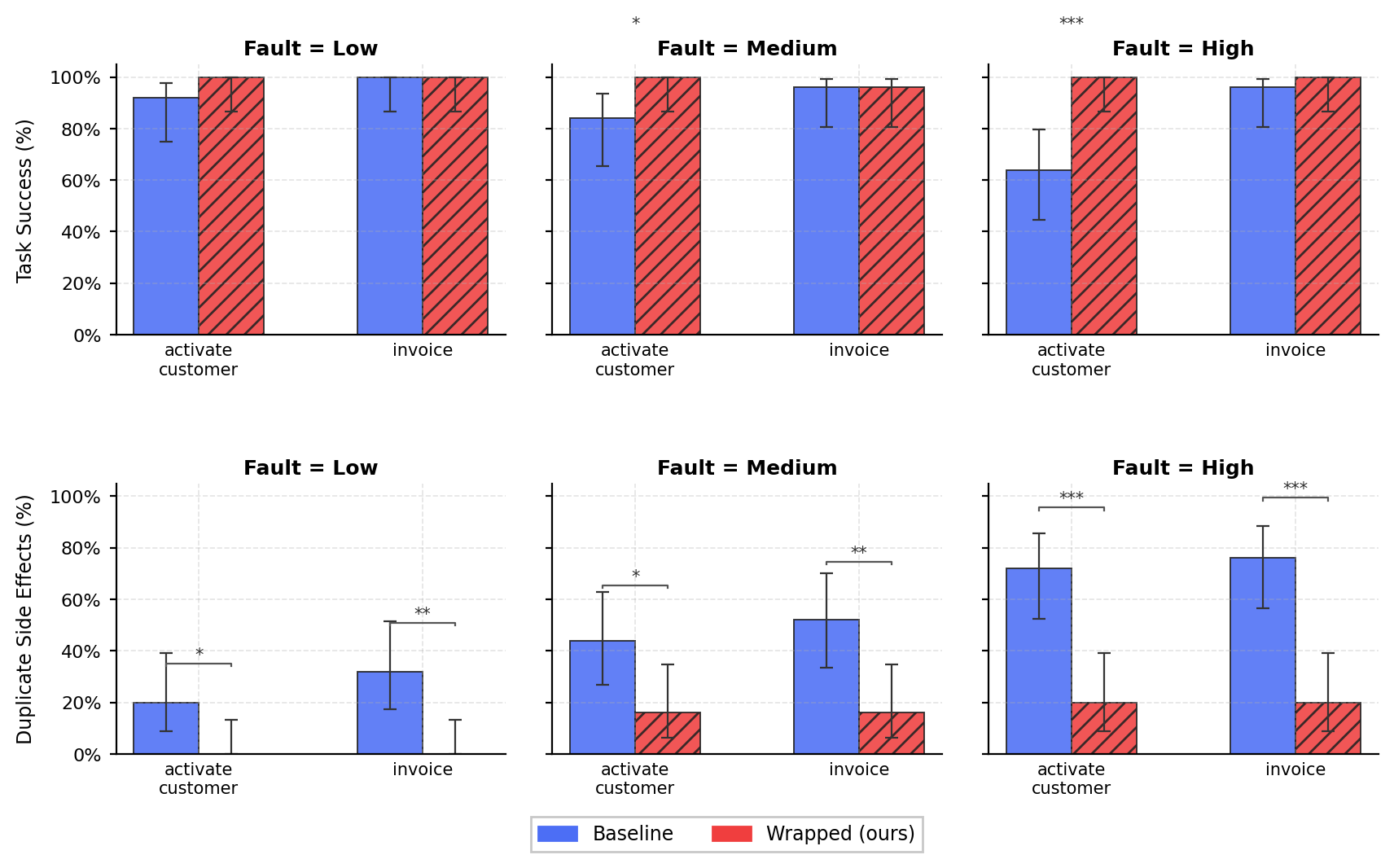}
\caption{Task success rate and duplicate side-effect rate for the baseline and verify-before-retry wrapped agent across fault levels.}
\label{fig:main-comparison}
\end{figure}

As shown in Figure ~\ref{fig:main-comparison}, overall, the wrapped agent performs better than the baseline, especially as the fault level increases. For the \texttt{activate\_customer} task, the wrapped agent achieved 100\% task success across all fault levels. In contrast, the baseline dropped from 92\% success at low fault to 64\% success at high fault. The largest improvement appears in the high fault setting, where the wrapped agent improves task success by 36 percentage points. Moreover, for the \texttt{invoice} task, both methods already perform well on task success. The baseline achieves 100\%, 96\%, and 96\% success across low, medium, and high fault levels, while the wrapped agent achieves 100\%, 96\%, and 100\%. This suggests that the wrapper is most useful when the task has a high risk of unsafe duplicate actions, such as sending messages.

\subsection{Verification Sharply Cuts Duplicate Side Effects}

Moreover, Figure ~\ref{fig:main-comparison} also shows a clear improvement in duplicate call side effects. The retry-only baseline frequently repeats actions after ambiguous failures, while the wrapped agent reduces duplicates by checking whether the action has already completed before retrying. On the \textbf{activate\_customer} task, duplicate side effects increase for the baseline as faults become more severe: 20\% at low fault, 44\% at medium fault, and 72\% at high fault. The wrapped agent reduces these to 0\%, 16\%, and 20\%, respectively.

The same pattern appears in the \textbf{record\_invoice} task. The baseline produces duplicate side effects in 32\%, 52\%, and 76\% of runs across low, medium, and high fault levels. The wrapped agent reduces these to 0\%, 16\%, and 20\%. The largest duplicate reductions, at high fault on both tasks and at low fault on invoice, are statistically significant, as shown in Figure ~\ref{fig:table_main}.

\subsection{Verification, Not Retry, Drives the Gains}

\begin{figure}
    \centering
    \includegraphics[width=0.9\linewidth]{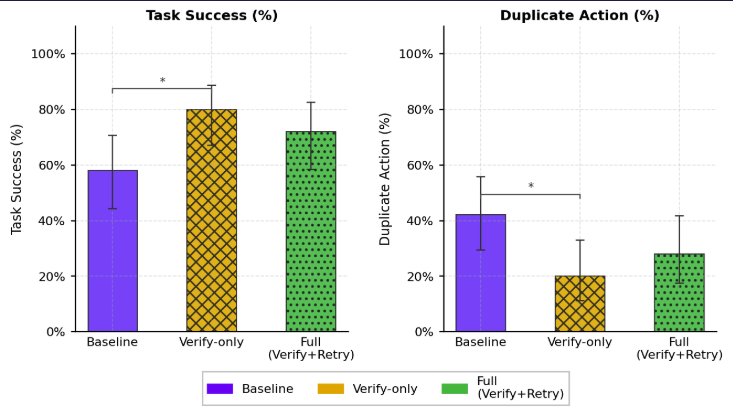}
    \caption{Graphic illustrating results from the ablation study to pin down which attribute of the verify-before-retry actually helped with the improvement from baseline.}
    \label{fig:ablation}
\end{figure}

To understand which part of the wrapper was responsible for the improvement, we compared three versions of the agent: Baseline, verify-only, and verify-before-retry. The baseline retries tool calls after an error without checking whether the action already happened. The verify-only version checks the external state after an uncertain tool call but does not retry. The Full version checks first and retries only when verification suggests the action did not complete.

As shown in Figure ~\ref{fig:ablation}, verification is the main factor driving the improvement. The baseline achieved about 58\% task success, while verify-only increased success to about 80\%. It also reduced duplicate actions from about 42\% to 20\%. The Full method still improved over the baseline, reaching about 72\% task success and reducing duplicate actions to about 28\%.

These rates come from a separate ablation run and should be read relative to one another, not against the absolute values in Figures 2 and 4. Verify-only performed comparably to the verify-before-retry version in this run. This suggests that retries are not always beneficial: while retrying can fix real failures, it also creates another chance for duplicated actions or new tool faults. Overall, the ablation shows that checking whether the action already happened before retrying is the most important part of the system, and retries should be used carefully rather than automatically.

\subsection{Summary}

\begin{figure}[h]
\centering
\includegraphics[width=1\linewidth]{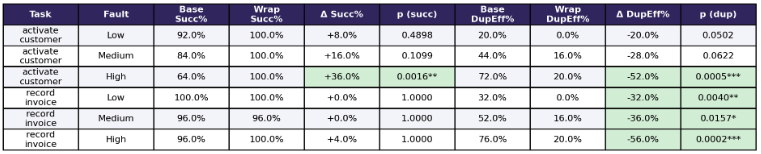}
\caption{Comparison of task success and duplicate side effect rates between baseline and wrapper.}
\label{fig:table_main}
\end{figure}

The results, as shown in Figure~\ref {fig:table_main}, support our central claim: a tool-using agent should not blindly retry after an ambiguous tool failure. The retry-only baseline often repeats actions that already succeeded, especially under higher fault rates. The verify-before-retry wrapper reduces this problem by checking the external state before authorizing a retry. 


The benefit is most visible in duplicate side effect reduction, which is significant across all tested settings. Task success also improves strongly for \textbf{activate\_customer}, where duplicate messages directly violate the task objective. For \textbf{record\_invoice}, task success is already high, but the wrapper still improves execution safety by reducing duplicate writes. Therefore, these results show that reliable tool use evaluation should measure not only whether the final state is correct, but also whether the agent reached that state safely.

\section{Discussion}

Our results show that verify-before-retry improves the reliability of LLM agents operating under non-atomic tool failures. The largest gains come from preventing unnecessary retries after ambiguous failures, leading to substantially fewer duplicate side effects while maintaining high task success across increasing fault levels. This effect is particularly evident in workflows where duplicate actions are themselves correctness violations. Still, it also extends to tasks where the final state may appear correct despite redundant external writes. These findings suggest that evaluating only final task success can obscure important reliability failures, since an agent may eventually reach the correct outcome through an unsafe execution path.

More broadly, these results reinforce a principle long recognized in distributed systems: an error response does not necessarily imply that an operation failed. By verifying postconditions before retrying, the agent bases recovery decisions on the external system state rather than on potentially misleading execution signals. This suggests that future agent frameworks and tool APIs should provide stronger execution semantics, such as idempotency keys, operation identifiers, or explicit status endpoints, enabling agents to recover safely from ambiguous failures without requiring changes to the underlying language model.

Our study has several limitations. The evaluation is conducted in a controlled simulator with two representative workflows and hand-designed postcondition verifiers, rather than production APIs with more diverse behaviors. Consequently, the reported improvements may not capture all real-world failure modes, including stale reads, authentication failures, or incomplete verification logic. Furthermore, the wrapper cannot eliminate failures when the verifier itself observes outdated or incomplete state. Despite these limitations, our results demonstrate that introducing explicit verification between failure detection and retry is a simple yet effective mechanism for improving the reliability and safety of tool-using LLM agents.

\section{Future Work}
    
This study shows that verifying the external state before retrying a tool call can make LLM agents more reliable under non-atomic tool failures. Our results suggest several directions for extending verify-before-retry. First, the current framework makes binary verification decisions, treating an operation as either complete or incomplete. A promising direction is to replace this decision with a confidence-based estimate that combines multiple signals, such as postcondition verification, tool responses, idempotency information, and execution history. Rather than always retrying after a failed verification, an agent could use this confidence to decide whether to retry immediately, wait and verify again, request human intervention, or terminate safely.

Second, the reliability of the framework depends on the quality of the postcondition verifier. While our experiments use task-specific verifiers, more complex workflows require richer semantic checks that validate the completeness and correctness of external state rather than simple object existence. Developing general methods for automatically constructing or learning postcondition verifiers would substantially improve the practicality of the approach.

Finally, our evaluation was conducted in a controlled simulator with two representative workflows. Future work should evaluate verify-before-retry on real-world APIs and broader agent benchmarks involving databases, email systems, spreadsheets, and web services, where additional challenges such as eventual consistency, authentication failures, rate limiting, and evolving APIs arise. Such studies would establish the extent to which verification-based recovery generalizes across diverse models, tasks, and deployment environments.

\section{Conclusion}

This paper studied a common but often overlooked failure mode in LLM agents: non-atomic tool execution. In these cases, a tool action may actually be applied to the external world, but the agent receives an error, timeout, stale read, or incomplete response. When the agent retries without checking the true state of the world, it can create duplicate actions, incorrect final reports, or false success states.

To address this problem, we introduced a lightweight verify-before-retry wrapper. Instead of immediately retrying after an uncertain tool result, the wrapper first checks whether the intended postcondition has already been satisfied. If the action has already succeeded, the agent avoids an unnecessary retry. If the action has not succeeded, the agent can retry with better evidence. Our results show that this simple system-level change can improve reliability without requiring any changes to the underlying LLM.

Overall, this work shows that reliable LLM agents cannot depend only on model reasoning ability. Even strong models can fail when the surrounding tool environment is ambiguous or non-atomic. Building safer agents therefore requires external verification, careful retry policies, and explicit checks on the world state. Verify-before-retry is a practical step toward more dependable LLM agents, especially in workflows where duplicate or incorrect actions can have real consequences.

\bibliographystyle{unsrt}  





\bibliography{references}

\end{document}